\documentclass[prd,showpacs,showkeys,floatfix,nofootinbib,eqsecnum,
               preprint,12pt,tightenlines,fleqn]{revtex4-1} 
\usepackage{amssymb,epsf}
\usepackage{latexsym,,amssymb}
\usepackage{amsmath,mathrsfs}
\usepackage{graphics,epsfig}
\usepackage[english]{babel}

\newcommand{\beq}{\begin{equation}}
\newcommand{\eeq}{\end{equation}}
\newcommand{\beqa}{\begin{eqnarray}}
\newcommand{\eeqa}{\end{eqnarray}}
\newcommand{\bsubeqs}{\begin{subequations}}
\newcommand{\esubeqs}{\end{subequations}}

\begin{document}
%
%
\noindent Phys. Rev. D 89, 084064 (2014) 
\hfill  arXiv:1303.7219\newline\vspace*{2mm}
\title{\vspace*{2mm}A nonsingular spacetime defect\\[2mm]\vspace*{2mm}}
\author{F.R.~Klinkhamer}
\email{frans.klinkhamer@kit.edu}
\author{C.~Rahmede}
\email{christoph.rahmede@kit.edu}\affiliation{Institute for
Theoretical Physics, Karlsruhe Institute of
Technology (KIT), 76128 Karlsruhe, Germany\\}

\begin{abstract}%
\noindent 
A nonsingular localized static classical solution
is constructed for standard Einstein gravity
coupled to an $SO(3)\times SO(3)$ chiral model
of scalars (Skyrme model).
This solution corresponds to a spacetime defect
and its construction proceeds in three steps.
First, an \textit{Ansatz} is presented for a
solution with nonsimply connected topology of the spacetime manifold.
Second, an exact vacuum solution of the reduced field equations
is obtained.
Third, matter fields are included and a particular
exact solution of the reduced field equations is found.
The latter solution has a diverging total energy,
but its existence at least demonstrates 
that a nonsingular defect-type solution having 
nonsimply connected topology is possible with nontrivial matter fields.
\end{abstract}


\maketitle

\section{Introduction}
\label{sec:Introduction}

It can be argued~\cite{Wheeler1957,Wheeler1968}
on general grounds that the small-length-scale
structure of quantum spacetime is nontrivial.
This structure has been called a quantum spacetime foam.
Over larger length scales (lower energy scales),
an effective classical spacetime
manifold emerges and the crucial question is whether that
effective spacetime  is perfectly smooth or not.

Particle/wave propagation over Swiss-cheese-type manifolds  
has been studied and the problem  can, in principle,
be solved exactly~\cite{Bethe44,BernadotteKlinkhamer2006}.
The simplest example of a Swiss-cheese-type manifold  
has identical static ``defects'' (alternatively called ``holes'' or ``knots''),
where each defect provides nontrivial topology.
The particular defect considered in Ref.~\cite{BernadotteKlinkhamer2006}
has, however, a divergent (delta-function-type) Ricci curvature scalar
and does not solve the vacuum Einstein equations.

The goal, now, is to construct a nonsingular
defect solution by the use of appropriate coordinates and matter fields.
For the defect topology at hand
(holes with antipodal points identified; see below),
it has been suggested~\cite{Schwarz2010}
to use the gravitating $SO(3)$ Skyrme
model~\cite{Skyrme1961,Glendenning-etal1988,DrozHeuslerStraumann1991,%
BizonChmaj1992,HeuslerStraumannZhou1993,KleihausKunzSood1995}
with an additional interaction term~\cite{PottingerRathske1986}
giving negative energy-density contributions.
The motivation for using the $SO(3)$ Skyrme field is to allow 
for the possibility of having a topologically stable solution consistent with the
boundary conditions at the defect core.
Negative energy-density contributions may turn out
to be essential for a satisfactory defect solution
over a nonsimply connected spacetime~\cite{Gannon1975,Klinkhamer2013-JETPL}.
For this reason, our analysis allows for the possibility
of having negative energy-density contributions, 
even though,  at this stage, we can do without.  

The present article uses a new \textit{Ansatz} for these
fields and gives special attention to the
behavior of the reduced field equations
at the defect core. We are then able to obtain a nonsingular
defect solution in the gravitating $SO(3)$ Skyrme model.

The main results of this article are, first, an exact vacuum
solution and, second, an exact nonvacuum solution. 
Physically, the vacuum solution will be relevant far away 
from localized energy-momentum distributions.
The importance of the particular nonvacuum solution found
is that it shows how the matter is distributed, 
given that the topology is the  same as that of
the vacuum solution.

\section{Manifold}
\label{sec:Manifold}

The four-dimensional spacetime manifold considered in this article 
has topology
\beq\label{eq:M4}
M_4 = \mathbb{R} \times M_3\,.
\eeq
The nontrivial topology appears in the 3-space $M_3$,
which is, in fact,  a noncompact,
orientable, nonsimply connected manifold without boundary.
Up to a point, $M_3$ is homeomorphic
to the three-dimensional real-projective space,
\beq\label{eq:M3-topology}
M_3 \simeq
\mathbb{R}P^3 - \{\text{point}\}\,.
\eeq
Further details can be found in
Refs.~\cite{BernadotteKlinkhamer2006,Schwarz2010}.
Here, only the absolutely necessary information will be given.

\begin{figure*}[b] 
\vspace*{2mm}
\includegraphics[width=0.45\textwidth]{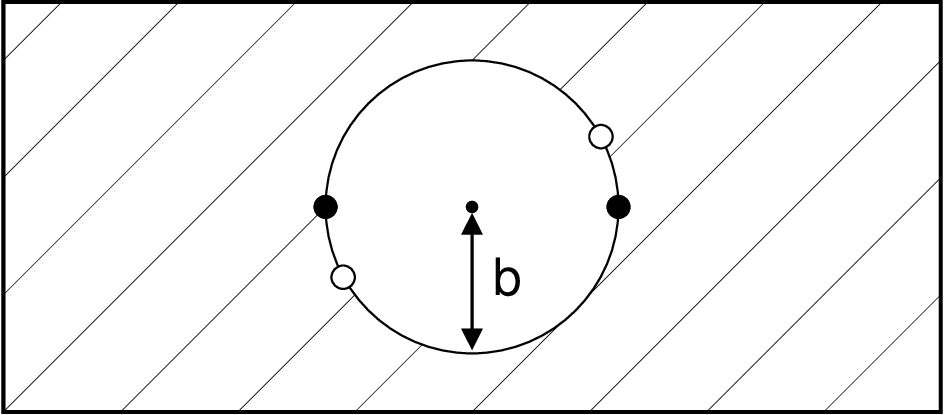}
\caption{Three-space $M_3$ obtained by surgery on $\mathbb{R}^3\,$:
the interior of the ball with radius $b$ is removed
and antipodal points on the boundary of the ball are
identified (as indicated by open and filled circles).}
\label{fig:defect}
\end{figure*}

For the explicit construction of $M_3$, we perform local
surgery on  the three-dimensional Euclidean space
$E_3=\big(\mathbb{R}^3,\, \delta_{mn}\big)$.
We use the standard Cartesian and spherical
coordinates on $\mathbb{R}^3$,
\beq\label{eq:Cartesian-spherical-coord}
\vec{x}
\equiv |\vec{x}|\, \widehat{x}
= (x^1,\,  x^2,\, x^3)
= (r \sin\theta  \cos\phi,\,r \sin\theta  \sin\phi,\, r \cos\theta )\,,
\eeq
with $x^m \in (-\infty,\,+\infty)$,
$r \geq 0$, $\theta \in [0,\,\pi]$, and $\phi \in [0,\,2\pi)$.
We now obtain $M_3$  from $\mathbb{R}^3$ by removing
the interior of the ball $B_b$ with radius $b$ and
identifying antipodal points on the boundary $S_b \equiv \partial B_b$.
With point reflection denoted by $P(\vec{x})=-\vec{x}$,
the 3-space $M_3$ is given by
\beq\label{eq:M3-definition}
M_3 =
\big\{  \vec{x}\in \mathbb{R}^3\,:\; \big(|\vec{x}| \geq b >0\big)
\wedge
\big(P(\vec{x})\cong \vec{x} \;\;\text{for}\;\;  |\vec{x}|=b\big)
\big\}\,,
\eeq
where $\cong$ stands for point-wise identification (Fig.~\ref{fig:defect}).

The single set of coordinates \eqref{eq:Cartesian-spherical-coord}
does not suffice for an appropriate description of $M_3$.
The reason is simply that two different values of these coordinates
may correspond to a single point. For example,
$\vec{x}=(b,\,0,\,0)$ and $\vec{x}=(-b,\,0,\,0)$
correspond to the same point of $M_3$.
A relatively simple covering of $M_3$ uses three sets of
coordinates (also called charts or patches),
labeled by $n=1,2,3$. Each coordinate chart surrounds
one of the three Cartesian coordinate axes.
These coordinates are denoted by
\beq\label{eq:XnYnZn}
(X_n,\,  Y_n,\, Z_n)\,, \quad\text{for}\quad n=1,\,2,\,3\,,
\eeq
and are, despite appearances, triples of non-Cartesian coordinates.
Specifically, the set of coordinates surrounding the
$x^2$-axis segment with $|x^2|\geq b$ is given by
\bsubeqs\label{eq:X2Y2Z2-def}
\beqa
X_2 &=& \left\{\begin{array}{ll}
                \phi       &\quad\text{for}\quad  0 < \phi< \pi\,,\\
                \phi-\pi   &\quad\text{for}\quad  \pi < \phi < 2\pi\,,
              \end{array}\right.\\[2mm]
Y_2 &=& \left\{\begin{array}{ll}
                r-b       &\quad\text{for}\quad  0 < \phi< \pi\,,\\
                b-r       &\quad\text{for}\quad  \pi < \phi < 2\pi\,,
              \end{array}\right.\\[2mm]
Z_2 &=& \left\{\begin{array}{ll}
                \theta       &\quad\text{for}\quad  0 < \phi< \pi\,,\\
                \pi-\theta   &\quad\text{for}\quad  \pi < \phi < 2\pi\,,
              \end{array}\right.
\eeqa
\esubeqs
with ranges
\bsubeqs\label{eq:X2Y2Z2-ranges}
\beqa
X_2 &\in& (0,\,\pi)\,,\\[2mm]
Y_2 &\in& (-\infty,\,\infty)\,,\\[2mm]
Z_2 &\in& (0,\,\pi)\,.
\eeqa
\esubeqs
The other two sets, $(X_1,\,  Y_1,\, Z_1)$ and $(X_3,\,  Y_3,\, Z_3)$,
are defined similarly.

In the following, we consider spherically symmetric fields
and it suffices to consider one coordinate chart,
which we take to be \eqref{eq:X2Y2Z2-def}.
The notation is, furthermore, simplified as follows:
\beq\label{eq:XYZ-def}
(X,\,  Y,\, Z,\,T) \equiv
(X_2,\,  Y_2,\, Z_2,\,T)\,,
\eeq
where the time coordinate has been added in order
to describe the spacetime manifold $M_4$.

\section{Fields and action}
\label{sec:Fields-and-action}

The spacetime manifold \eqref{eq:M4} of the previous section is now
equipped with a metric $g_{\mu\nu}(X)$, whose dynamics are
governed by the standard Einstein--Hilbert
action~\cite{Wald1984}. In addition, there is a
scalar field $\Omega(X)\in SO(3)$, with self-interactions
determined by a quartic Skyrme term in the matter action~\cite{Skyrme1961}
and by another quartic term~\cite{PottingerRathske1986} whose
coupling constant $\gamma$ is taken to be non-negative,
allowing for negative energy-density contributions.

Specifically, the combined action of the pure-gravity sector,
labeled ``grav,'' and the matter sector, labeled ``mat,'' 
is given by ($c=\hbar=1$)
\bsubeqs\label{eq:action}
\begin{equation}\label{eq:action-S}
 S=\int_{M_4} d^4X\,\sqrt{-g}\,
 \Big(\mathcal{L}_\text{grav,\,EH}+\mathcal{L}_\text{mat,\,kin}
 +\mathcal{L}_\text{mat,\,Skyrme}+ \mathcal{L}_{\text{mat,\,metastab}}\Big)\,,
\end{equation}
with Lagrange densities
\begin{eqnarray}
 \mathcal{L}_\text{grav,\,EH}&=&
\frac{1}{16\pi G_N}\:R\,,
 \\[2mm]
 \mathcal{L}_\text{mat,\,kin}&=&
\frac{f^2}{4}\:\text{tr}\big(\omega_\mu\,\omega^\mu\big)\,,
 \\[2mm]
 \mathcal{L}_\text{mat,\,Skyrme}&=&
\frac{1}{16 e^2}\: \text{tr}\Big(\left[\omega_\mu,\,\omega_\nu\right]
\left[\omega^\mu,\,\omega^\nu\right]\Big)\,,\\[2mm]
 \mathcal{L}_{\text{mat,\,metastab}}&=&
\gamma\,\frac{1}{48e^2}\;\Big( \text{tr}(\omega_\mu\,\omega^\mu)\Big)^2\,,
\end{eqnarray}
\esubeqs
in terms of the Ricci curvature scalar $R$ and
\begin{equation}
\omega_\mu \equiv \Omega^{-1}\,\partial_\mu\,\Omega\,.
\end{equation}
The $SO(3)\times SO(3)$ global symmetry of the matter sector
is realized on the scalar field by the following transformation
with constant parameters $S_L,\, S_R \in SO(3)$:
\beq
 \Omega(X) \to S_L \cdot \Omega(X) \cdot S_R^{-1}\,,
\eeq
where the central dot denotes matrix multiplication.
As mentioned in the last paragraph of Sec.~\ref{sec:Manifold},
the generic argument $X$ of the fields and
the measure $d^4 X$ in the integral \eqref{eq:action-S}
correspond to only one of the
three coordinate charts needed to cover $M_4$.

\section{Ansatz and field equations}
\label{sec:Ansatz-and-field-equations}

\subsection{Ansatz}\label{subsec:Ansatz}

A spherically symmetric \textit{Ansatz}
for the metric is given by the following line element:
\bsubeqs\label{eq:metric-Ansatz-W-definition}
\beqa\label{eq:metric-Ansatz}
 ds^2 &=&-
\exp\big[2\,\widetilde{\nu}(W)\big]\, dT^2
 +
\exp\big[2\,\widetilde{\lambda}(W)\big]\, dY^2
 +W \left(dZ^2+\sin^2 Z\, dX^2\right)\,,
\\[2mm]
\label{eq:W-definition}
W &\equiv& b^2+Y^2\,,
\eeqa
\esubeqs
where, as explained at the end of Sec.~\ref{sec:Manifold},
we only show the coordinates of one chart
with $Y \in (-\infty,\,\infty)$.
For later convenience, two further functions are introduced:
\bsubeqs\label{eq:kappatilde-mutilde}
\begin{eqnarray}\label{eq:kappatilde}
\widetilde{\kappa}(W) 
&\equiv& 
\exp\big[\widetilde{\lambda}(W)\big]\,,
\\[2mm]
\label{eq:mutilde}
\widetilde{\mu}(W) 
&\equiv& 
\exp\big[\widetilde{\nu}(W)\big]\,.
\end{eqnarray}
\esubeqs
The scalar field is given by the hedgehog-type
\textit{Ansatz}~\cite{Schwarz2010,Skyrme1961},  
\bsubeqs\label{eq:hedgehog-Ansatz}
\begin{eqnarray}\label{eq:hedgehog-Ansatz-Omega}
\Omega &=&
\cos\big[\widetilde{F}\left(r^2\right)\big]\;\openone
-\sin\big[\widetilde{F}\left(r^2\right)\big]\;
\widehat{x}\cdot \vec{S}
+\big(1-\cos\big[\widetilde{F}\left(r^2\right)\big]\big)\;
\widehat{x} \otimes \widehat{x}\,,
\\[2mm]\label{eq:hedgehog-Ansatz-bcs}
\widetilde{F}(b^2) &=& \pi\,,
\\[2mm]
S_1 &\equiv&  \left(
                \begin{array}{ccc}
                  0 & 0 & 0 \\
                  0 & 0 & 1 \\
                  0 & -1 & 0 \\
                \end{array}
              \right)\,,
\quad
S_2 \equiv  \left(
                \begin{array}{ccc}
                  0 & 0 & -1 \\
                  0 & 0 & 0 \\
                  1 & 0 & 0 \\
                \end{array}
              \right)\,,
\quad
S_3 \equiv  \left(
                \begin{array}{ccc}
                  0 & 1 & 0 \\
                  -1 & 0 & 0 \\
                  0 & 0 & 0 \\
                \end{array}
              \right)\,,
\end{eqnarray}
\esubeqs
with $(\widehat{x}\otimes \widehat{x})^{ab}= \widehat{x}^a\, \widehat{x}^b$
in components.
Note that, because of the boundary condition
\eqref{eq:hedgehog-Ansatz-bcs} at $|\vec{x}|=b$,
it is possible to use the single coordinate
chart \eqref{eq:Cartesian-spherical-coord}
with the further identification
$ r^2 = b^2+Y^2 = W $ for the coordinates
used in the metric \eqref{eq:metric-Ansatz-W-definition}.
In other words, the topology of $M_3$ (see Fig.~\ref{fig:defect})
is trivially consistent with the
hedgehog field having the boundary value $\widetilde{F}(b^2)=\pi$.

The arguments of the tilde functions in the
above \textit{Ans\"{a}tze} have dimension length-square.
Related functions with lengths as arguments
can be defined as follows:
\beq
\nu\left(\sqrt{b^2+Y^2}\right)
\equiv
\widetilde{\nu}\left(b^2+Y^2\right)\,,
\eeq
and similarly for $\lambda$ and $F$.
The functions without the tilde resemble those of the previous
literature~\cite{Skyrme1961,Glendenning-etal1988,DrozHeuslerStraumann1991,%
BizonChmaj1992,HeuslerStraumannZhou1993,KleihausKunzSood1995,%
PottingerRathske1986},
but we prefer to work with the tilde functions.

The relevant nonvanishing components of the Riemann tensor for the
\textit{Ansatz} \eqref{eq:metric-Ansatz-W-definition} are
\bsubeqs\label{eq:Riemann}
\begin{eqnarray}
R^T_{\ Y T Y}&=&-2 \left[\widetilde{\nu}'
+2 Y^2\left(- \widetilde{\nu}'\,\widetilde{\kappa}'/\widetilde{\kappa}
+ \widetilde{\nu} ''+ \widetilde{\nu} '^2\right)\right]\,, \\[1mm]
R^T_{\ Z T Z}&=&-2 Y^2\,  \widetilde{\nu}'/\widetilde{\kappa}^{2}\,, \\[1mm]
R^T_{\ X T X}&=&
\sin ^2 Z \, R^T_{\ Z T Z}\,,
\\[1mm]
R^Y_{\ Z Y Z}&=&\frac{2 Y^2  \left(b^2+Y^2\right)
\widetilde{\kappa}'/\widetilde{\kappa}-b^2}{(b^2+Y^2)\,{\widetilde{\kappa}^{2}}}\,, \\[1mm]
R^Y_{\ X Y X}&=&\sin ^2 Z\, R^Y_{\ Z Y Z}\,,
\\[1mm]
R^Z_{\ X Z X}&=& \sin ^2 Z\,\left(1-\frac{Y^2}{(b^2+Y^2)\,{\widetilde{\kappa}^{2}}}\right)\,,
\end{eqnarray}
\esubeqs
where the prime stands for differentiation with respect to $W$.
The components not shown in \eqref{eq:Riemann} either vanish
or can be computed from the ones above by the usual symmetry properties.

\subsection{Reduced field equations}
\label{subsec:Reduced-field-equations}

The derivation of the reduced field equations is
straightforward~\cite{Endnote:var-eqs}. Henceforth, we will use
the following dimensionless model parameters and dimensionless  variables:
\bsubeqs\label{eq:dimensionless}
\begin{eqnarray}\label{eq:dimensionless-eta}
\widetilde{\eta}&\equiv&8\pi\eta \equiv 8\pi\, G_N\, f^2\,,\\[2mm]
w  &\equiv& (e\,f)^2\;W = (y_0)^2+y^2\,,\\[2mm]
y  &\equiv& e\,f\;Y\,,\\[2mm]
y_0 &\equiv& e\,f\;b\,.  
\end{eqnarray}
\esubeqs

The reduced Einstein gravitational field equations and reduced 
matter field equations will be given 
in Appendix~\ref{app:ODEs}. From these equations, 
one obtains three ordinary differential equations (ODEs):
\bsubeqs\label{eq:final-ODEs-short}
\begin{eqnarray}
\label{eq:final-ODEs-short-kappatilde}
 \widetilde{\kappa}'(w)&=&
\widetilde\kappa\left(
\frac{w(1-\widetilde\kappa^2)+y_0^2}{4w \left(w- y_0^2\right)}\right)
+ \widetilde{\eta}\,
Q_1\big[\widetilde{F}(w),\,\widetilde{\kappa}(w),\,\widetilde{\nu}(w)\big]\,,
\\[2mm]
\label{eq:final-ODEs-short-nutilde}
\widetilde{\nu}'(w)&=&
\frac{\widetilde{\kappa}(w)^2}{4 \left(w-y_0^2\right)}
-\frac{1}{4 w} + \widetilde{\eta}\,
Q_2\big[\widetilde{F}(w),\,\widetilde{\kappa}(w),\,\widetilde{\nu}(w)\big] \,,
\\[2mm]
\widetilde{F}''(w)&=&
Q_3[\widetilde{F}(w),\,\widetilde{\kappa}(w),\,\widetilde{\nu}(w)]\,,
\end{eqnarray}
\esubeqs
where the prime now stands for differentiation with respect to $w$ and
the three $Q_n$ are certain functionals given in Appendix~\ref{app:ODEs}.
Specifically, the functionals $Q_1$ and $Q_2$  
are given by the parts proportional to $\widetilde \eta$ 
in \eqref{eq:final-ODEs-a} and \eqref{eq:final-ODEs-b},
while $Q_3$ is given by the right-hand side of \eqref{eq:final-ODEs-c}.

The ODEs \eqref{eq:final-ODEs-short}
are to be solved with the following boundary conditions
for the nonvacuum solution:
\bsubeqs\label{eq:BCS}
\begin{eqnarray}\label{eq:BCS-F}
\widetilde{F}(y_{0}^2)&=& \widetilde{F}(\infty) = \pi\,,\\[2mm]
\label{eq:BCS-kappatilde-nutilde}
\widetilde{\kappa}(y_{0}^2)&=& 0\,,
\quad
\widetilde{\nu}(\infty)= 0\,.
\end{eqnarray}
\esubeqs
Physically, the reason for choosing these boundary conditions is to obtain 
local asymptotic flatness and to have a nonvanishing 
matter contribution that allows for an analytic solution.
For the vacuum solution, \eqref{eq:BCS-F} is  
to be replaced by
$\widetilde{F}(y_{0}^2)$ $=$ $\widetilde{F}(\infty)$ $=$ $0$.

The field configuration \eqref{eq:hedgehog-Ansatz}
with boundary conditions \eqref{eq:BCS-F} has a vanishing winding number 
[recall \eqref{eq:M3-topology} and 
the further homeomorphism $\mathbb{R}P^3 \simeq SO(3)$].   
It is, therefore, not a genuine Skyrmion
[which has $F(0) = \pi$ and $F(\infty) = 0$ for a unit winding number].  
Our matter solution may very well turn out to be unstable, 
but that is not important for the main goal of this article,
i.e., finding a nontrivial solution with matter fields
(see Sec.~\ref{sec:Discussion} for further discussion).

\section{Vacuum solution}
\label{sec:Vacuum-solution}

For vacuum matter fields,
\bsubeqs\label{eq:vacuum-solution}
\beqa
\widetilde{F}(w) &=& 0\,,
\eeqa
the ODEs \eqref{eq:final-ODEs-short-kappatilde}
and \eqref{eq:final-ODEs-short-nutilde}
have $Q_1=Q_2=0$ and can be solved exactly
for boundary conditions \eqref{eq:BCS-kappatilde-nutilde},
\beqa
\label{eq:vacuum-solution-kappatilde}
\widetilde{\kappa}(w) &\equiv& \exp\big[\widetilde{\lambda}(w)\big]
=
\sqrt{\frac{1-(y_0)^2/w}{1  -\ell/\sqrt{w}}}\;,
\;\\[2mm]
\label{eq:vacuum-solution-mutilde}
\widetilde{\mu}(w) &\equiv& \exp\big[\widetilde{\nu}(w)\big]
=
\sqrt{1-\ell/\sqrt{w}}\;,
\eeqa
\esubeqs
with a dimensionless constant $\ell$
and the definition $w \equiv y^2 +(y_0)^2$ in terms
of the ``radial'' coordinate $y \in (-\infty,\,+\infty)$
and the defect parameter $y_0>0$
(the corresponding dimensional parameter $b$ is also nonzero and positive).
Incidentally, the same $\widetilde{\kappa}(w)$ and $\widetilde{\mu}(w)$
solutions are obtained for the
case $\widetilde{F}(w) = \pi$ and $\widetilde{\eta}=0$,
which corresponds to the setup of the boundary conditions \eqref{eq:BCS-F}
and vanishing Newton's constant, $G_N=0$
(see Sec.~\ref{sec:Nonvacuum-solution}  for details).

At this point, we can make four general remarks.
First, the vanishing of the metric component $\widetilde{\kappa}^{\,2}$
at $Y=0$ implies that the defect of Fig.~\ref{fig:defect}
at that point $Y=0$ and fixed time $T$ has zero physical extent 
along the radial direction
(i.e., $ds=0$ at $Y=0$ for $dT=dX=dZ=0$).  

Second,  the real $\widetilde{\kappa}$
and $\widetilde{\mu}$ functions from \eqref{eq:vacuum-solution}, 
extended to the two other coordinate charts, cover
the whole of the manifold $M_4$ for the following
parameters:
\beq\label{eq:vacuum-solution-regular}
\ell < y_0\,,
\eeq
where $y_0$ has been assumed positive. The topology of this
solution is $\mathbb{R} \times M_3$ with $M_3$ given by \eqref{eq:M3-topology}.

Third, the solution behaves asymptotically ($w \sim y^2 \to \infty$)
 as follows:
\bsubeqs\label{eq:vacuum-solution-asymptotic}
\beqa
\widetilde{\kappa}^2 &\sim& 1\big/\big(1 -\ell/|y|\big)
\,,\\[2mm]
\widetilde{\mu}^2 &\sim& \big(1 -\ell/|y|\big)\,,
\eeqa
\esubeqs
which is to be compared to the standard
Schwarzschild metric~\cite{Wald1984} with mass $M$,
having the respective components
$1/(1- 2 G_N M/r)$ and $(1- 2 G_N M/r)$.
Note that the vacuum solution \eqref{eq:vacuum-solution} with
$\ell<0$ would produce ``antigravity,'' namely,
a  point mass far away from the
defect core would not be attracted towards it but repulsed.

Fourth,  
with the effective radial coordinate 
$\zeta\equiv \sqrt{b^2+Y^2}$ for $b>0$,
the metric of Sec.~\ref{sec:Vacuum-solution}
takes \emph{precisely} the standard Schwarzschild
form for \emph{all} values $\zeta\in [b,\,\infty)$, 
in line with Birkhoff's theorem as noted in Ref.~\cite{Klinkhamer2013-MPLA}.
The crucial point, however, is that the proper description
of the topology of the manifold requires the coordinate
$Y \in (-\infty,\,+\infty)$.

For $\ell< y_0$, all Riemann-curvature-tensor components  
\eqref{eq:Riemann} are finite over the whole manifold,
also at the defect core, $y=0$ or $w=(y_0)^2$. Specifically, we find for
the relevant nonvanishing Riemann tensor components:
\bsubeqs
\begin{eqnarray}
R^T_{\ Y T Y}&=& (e^2f^2)\;\frac{\ell
\left(w-y_0^2\right)}{w^2 \left(\sqrt{w}-\ell\right)}\,,\\[2mm]
R^T_{\ Z T Z}&=& \frac{-\ell}{2\sqrt{w}}\,,\\[2mm]
R^Y_{\ Z Y Z}&=& \frac{-\ell}{2\sqrt{w}}\,,\\[2mm]
R^Z_{\ X Z X}&=& \sin ^2 Z\,\frac{\ell  }{\sqrt{w}}\,.
\end{eqnarray}
\esubeqs
More importantly, the Kretschmann scalar obtained by
contraction of the Riemann tensor with itself,
\begin{eqnarray}
K &\equiv& R_{\mu\nu\rho\sigma}R^{\mu\nu\rho\sigma}
=12\;e^4 f^4 \;\frac{\ell^2}{w^3}\,,
\end{eqnarray}
remains finite over the whole of the manifold $M_4$,
because $w \geq (y_0)^2 >0$. This behavior contrasts with
that of the Schwarzschild metric over $\mathbb{R}^4$,
for which $K$ diverges at the point $r=0$.

A brief discussion of the geodesics
from the metric \eqref{eq:metric-Ansatz-W-definition} with
functions \eqref{eq:vacuum-solution-kappatilde} and
\eqref{eq:vacuum-solution-mutilde}
is given in Appendix~\ref{app:Geodesics}.  
The main result of this appendix is the existence of
radial geodesics passing through the center, which
illustrates the difference between our spacetime
and that of the standard Schwarzschild 
solution (cf. the fourth general remark above). 
Three follow-up 
papers~\cite{Klinkhamer2013-MPLA,Klinkhamer2013-APPB,Klinkhamer2013-review}
describe the global structure of the new vacuum solution,
in particular for the black-hole case $0<y_0<\ell$. 
These follow-up papers also give  
further details of the exact solution \eqref{eq:vacuum-solution} 
at the defect core~\cite{Endnote:blemish}.

Note, finally, that the nonsingular flat-spacetime metric   
is given by
\eqref{eq:vacuum-solution-kappatilde} and
\eqref{eq:vacuum-solution-mutilde} with $\ell=0$.
(Electromagnetic wave propagation
over this flat spacetime can be calculated with the
methods of Refs.~\cite{Bethe44,BernadotteKlinkhamer2006}.)
The actual value of the free parameter $\ell$
in the metric from \eqref{eq:vacuum-solution}
will have to be determined by adding matter
fields (the same applies for the
determination of $M$ in the standard Schwarzschild solution).

\section{Nonvacuum solution}
\label{sec:Nonvacuum-solution} 

\subsection{General solution}
\label{subsec:General-solution}

For the case of the constant Skyrme function $\widetilde{F}(w)=\pi$, 
it is possible to solve the reduced field equations
\eqref{eq:final-ODEs} exactly.
The vacuum solution of Sec.~\ref{sec:Vacuum-solution}
also had a constant Skyrme function, $\widetilde{F}(w)=0$,
and we can give a combined discussion by introducing the
constant 
\beq
\mathcal{F}=\pm 1
\eeq
so that the constant Skyrme function is given by
$\widetilde{F}(w)=\arccos\mathcal{F}$.
A further definition introduces the rescaled function 
\beq
\widetilde\sigma(w)\equiv
\frac{1}{\sqrt{1-y_0^2/w}}\;\;\widetilde\kappa(w)\,,
\eeq
which will simplify certain expressions below.

With these constant Skyrme functions for $\mathcal{F}=\pm 1$,
the differential equation for $\widetilde\kappa(w)$ 
reduces to a differential equation of Bernoulli type 
that can be solved with standard methods, while the 
differential equation for $\widetilde\mu(w)$ can be solved directly 
after insertion of the solution for $\widetilde\kappa(w)$. 
For $\widetilde\kappa(w)$ vanishing identically, 
one obtains $\widetilde\mu(w)=(w_0/w)^{1/4}$ with constant $w_0$.

For the case of nonvanishing $\widetilde\kappa(w)$, one obtains
the following solutions for $\mathcal{F}=\pm 1$:
\bsubeqs\label{eq:matter-solution}
\begin{eqnarray}
\widetilde{F}(w)
&=&
\arccos\mathcal{F} 
\,,\\[2mm]
\widetilde\sigma(w) &=&
\Big(\,1+C_{1}/\sqrt{w}+2\widetilde{\eta}\,(1-\mathcal{F})
\Big[(1-4\gamma/3)/w-1 \,\Big]\Big)^{-1/2}\,, 
\\[2mm]
\widetilde\mu(w) &=&
C_{2} \;\, \frac{1}{\widetilde\sigma(w)} \,, 
\end{eqnarray}
\esubeqs
having fixed the overall signs of $\widetilde\sigma$ 
and $\widetilde\mu$. There exists a finite positive value of 
the  quartic scalar coupling constant $\gamma$ above which 
the nonvacuum solution does not exist and the same holds for          
the dimensionless gravitational coupling constant
$\widetilde{\eta}$ (see below).

In preparation for the subsequent discussion, we already 
give the expressions of two curvature invariants.
The Ricci scalars of the two solutions given
in \eqref{eq:matter-solution} read
\beq\label{eq:R-vacuum-solution-nonvacuum-solution}
R\,\big|_{\mathcal{F}=\pm 1}
=4\, e^2 f^2\, \widetilde{\eta} \,\big(1-\mathcal{F}\big)\big/w\,.
\eeq
Similarly,
the Kretschmann scalar of the vacuum solution \eqref{eq:matter-solution} 
is given by
\bsubeqs\label{eq:K-vacuum-solution-nonvacuum-solution}
\begin{eqnarray}\label{eq:K-vacuum-solution}
K\,\big|_{\mathcal{F}=+1} 
&=&
12 \; e^4 f^4 \; C_1^2/w^3
\end{eqnarray}
and the one of the nonvacuum solution \eqref{eq:matter-solution}  by
\begin{eqnarray}\label{eq:K-nonvacuum-solution}
K\,\big|_{\mathcal{F}=-1} 
&=& (4/9)\; e^4 f^4 \;
\Big(
27\,  C_1^2\,  w
-72\,  C_1\,  \widetilde{\eta} \, \sqrt{w}\,   [w+8 \gamma -6]
\nonumber\\
&&
+16\,  \widetilde{\eta}^2\,  
\left[9\,  w^2+6\,  w \, (4 \gamma-3)+14\,  (3-4 \gamma)^2\right]
\Big) \Big/w^4,
\end{eqnarray}
\esubeqs
where the dependence on $C_{2}$ has canceled out 
(see the remark in the next subsection).

\subsection{Boundary conditions and asymptotics}
\label{subsec:Asymptotics-boundary-conditions}

The solution \eqref{eq:matter-solution} 
of the ODEs \eqref{eq:final-ODEs} is fixed by the boundary
conditions on $\widetilde{F}(w)$, $\widetilde\kappa(w)$, and  $\widetilde\mu(w)$
at the defect position $w=y_0^2$.
For $\widetilde{F}(w)$, the boundary conditions are 
$\widetilde{F}(y_0^2)=\pi$ and $\widetilde{F}'(y_0^2)=0$.
The value of $\widetilde\sigma(w)$ at $w=y_0^2$
fixes the constant $C_{1}$ of the solution \eqref{eq:matter-solution}.
Subsequently, the value of $\widetilde\mu(w)$ at $w=y_0^2$
fixes the constant $C_{2}$.
We remark that the actual value of $C_{2}$ has no direct physical
significance, as it can be changed by a rescaling of
the coordinate $T$, according to
\eqref{eq:metric-Ansatz} and \eqref{eq:mutilde}.

As an example, consider this particular set of boundary conditions
\bsubeqs\label{eq:bcs-massless-defect}
\beqa\label{eq:bcs-massless-defect-F}
\widetilde{F}(y_0^2)
&=& 
\pi \,,\quad
\widetilde{F}'(y_0^2)= 0\,, 
\\[2mm]
\label{eq:bcs-massless-defect-sigma}
\widetilde\sigma(y_0^2)
&=&
\Big(1+4\,\widetilde{\eta}\,\Big[(1-4\gamma/3)/y_0^2-1 \Big]\Big)^{-1/2}
\,,
\\[2mm]
\label{eq:bcs-massless-defect-mu}
\widetilde\mu(y_0^2)
&=& 
\frac{1}{1-4 \,\widetilde{\eta}}\;\frac{1}{\widetilde\sigma(y_0^2)}\, 
\,,
\eeqa\esubeqs
which gives the following constants in 
the solution \eqref{eq:matter-solution}:
\bsubeqs\label{eq:constants-massless-defect}
\beqa
\mathcal{F}
&=&
-1\,,
\\[2mm]
C_{1}
&=&  
0\,,
\\[2mm]
C_{2}
&=&
\frac{1}{1-4 \,\widetilde{\eta}} \,. 
\eeqa\esubeqs
The solution with $C_{1}\ne 0$ is obtained      
by shifting the boundary value $\widetilde\sigma(y_0^2)$  
away from the value on the right-hand side 
of \eqref{eq:bcs-massless-defect-sigma}.

Turning towards the asymptotics at spatial infinity, 
the solution \eqref{eq:matter-solution} has the following behavior
for $w\to\infty$:
\bsubeqs\beqa
\widetilde{F}(w)
&\sim& 
\pi \,,
\\[2mm]
\widetilde\sigma(w)
&\sim&  
\left(1+C_{\infty}+C_{1}/\sqrt{w}\,\right)^{-1/2}\,,
\\[2mm]
\widetilde\mu(w)
&\sim&  
C_{2}\,\left(1+C_{\infty}+C_{1}/\sqrt{w}\,\right)^{1/2}\,,  
\eeqa
\esubeqs
with
\beqa\label{eq:C-infty}
C_{\infty}
&\equiv&
-2 \,\widetilde{\eta}\,(1-\mathcal{F}) 
= \left\{\begin{array}{cl}
              0  &\quad\text{for}\quad  \mathcal{F}=+1\,,\\
              -4 \,\widetilde{\eta}  &\quad\text{for}\quad  \mathcal{F}=-1\,.
              \end{array}\right.
\eeqa
The case $\widetilde{\eta} \geq 1/4$ will be excluded in the following.

With the choice
\beq\label{eq:Cs-special}
C_{2}=
\frac{1}{1+C_{\infty}}\,,  
\eeq
one has the asymptotic values
\bsubeqs\beqa
\widetilde{F}_{\infty}
&=& 
\pi \,,
\\[2mm]
\widetilde\kappa_{\infty}
&=&  
1/\sqrt{1+C_{\infty}}\,,
\\[2mm]
\widetilde\mu_{\infty}
&=& 
1/\sqrt{1+C_{\infty}}\,.
\eeqa\esubeqs
Rescaling the coordinates $T$ and $Y$ 
of the \textit{Ansatz} \eqref{eq:metric-Ansatz-W-definition} by the same factor,
\bsubeqs\label{eq:rescaling-T-Y}
\beqa
\widehat{T}
&=&  
T/\sqrt{1+C_{\infty}}\,,
\\[2mm]
\widehat{Y}
&=& 
Y/\sqrt{1+C_{\infty}} \,,
\eeqa\esubeqs
then asymptotically reproduces the vacuum-solution functions  
\eqref{eq:vacuum-solution-kappatilde}
and \eqref{eq:vacuum-solution-mutilde} with
\bsubeqs\label{eq:rescaling-ell-y0}
\beqa
\ell &=& - C_{1}/(1+C_{\infty})^{3/2}   
\eeqa
if $y_0$ is also rescaled,
\beqa
\widehat{y_0} 
&=& y_0/\sqrt{1+C_{\infty}}\,.
\eeqa\esubeqs

Three  remarks are in order.
First, the particular $C_{2}$ value \eqref{eq:Cs-special}  
and coordinate rescaling \eqref{eq:rescaling-T-Y}
make the solution \eqref{eq:matter-solution}
consistent with the previous boundary
condition on $\widetilde\nu(\infty)$ from \eqref{eq:BCS-kappatilde-nutilde}.
Second, for this $C_{2}$ value and coordinate rescaling,
the functional behavior of 
\eqref{eq:matter-solution} in the theory with coupling constant $\gamma=3/4$
is identical to that of the 
vacuum-solution functions \eqref{eq:vacuum-solution},
but the spacetimes are different  [the rescaling of $Y$ and $b$ 
changes the angular part of the line element \eqref{eq:metric-Ansatz}:
spherical symmetry is preserved, but the metric is not Minkowski  
in the limit $|\widehat{Y}| \to \infty$].  
Third, following up on the previous remark, the asymptotic   
spacetime of the nonvacuum solution
has a deficit solid angle by a factor of $(1-4 \,\widetilde{\eta})$,
which is similar to the deficit angle of a two-dimensional cone
(for another solution with
deficit solid angle, see Ref.~\cite{VilenkinShellard1986}).

Expanding on the last remarks, consider the curvature 
invariants given in the final paragraph of 
Sec.~\ref{subsec:General-solution}.
The curvature invariants \eqref{eq:R-vacuum-solution-nonvacuum-solution} 
and \eqref{eq:K-vacuum-solution-nonvacuum-solution}
vanish for $w\to\infty$, as do the other 12 independent curvature 
invariants~\cite{Haskins1902,GeheniauDebever1956,Witten1959,Greenberg1972} 
which have been calculated explicitly but shall not be displayed 
here~\cite{Endnote:invariants}. With $w \to \infty$,  
the nonvacuum solution \eqref{eq:matter-solution}  for $\mathcal{F}=-1$
does not follow the vacuum solution 
\eqref{eq:matter-solution} for $\mathcal{F}=+1$, as 
the nonvacuum solution has $K\propto 1/w^2$  for $\widetilde{\eta}\ne 0$,
according to \eqref{eq:K-nonvacuum-solution},
and the vacuum solution has $K\propto 1/w^3$,
according to \eqref{eq:K-vacuum-solution}.
This unusual behavior of the nonvacuum solution
is due to the relatively slow decrease of the energy-momentum 
tensor with increasing values of $w$ as will
be discussed in the next subsection.

\subsection{Energy-momentum tensor}
\label{subsec:Energy-momentum-tensor}

The energy-momentum tensor $T_{\mu\nu}$ is diagonal 
for the \textit{Ansatz} fields.
The dimensionless energy density takes the form
\begin{equation}\label{EM:t_tt}
 t_{tt}\equiv 
T_{TT}/(e^2\,f^4)=
 \frac{2}{w}\,C_{2}^2\,   
\Big[1+\mathcal{A}(w)\Big]\, \Big(1-\mathcal{F}\Big) \,
\Big[1+C_{1}/\sqrt{w}-2 \widetilde{\eta}\,
\big(1-\mathcal{A}(w)\big)\, (1-\mathcal{F}) \Big]\,,
\end{equation}
with
\beq
\mathcal{A}(w)\equiv\left(1-4\gamma/3\right)\frac{ (1-\mathcal{F})}{2 w}\,.
\eeq 
For the nonvacuum case $\mathcal{F}=-1$, 
this gives a diverging total energy when integrated over 
the whole space (the volume integral includes $w$-integration over terms 
that asymptotically go as $w^{-1/2}$). One has $t_{tt}<0$ 
if one of the square brackets in (\ref{EM:t_tt}) becomes negative, but not both. 

The dimensionless radial pressure component is given by  
\begin{equation} 
 t_{yy}\equiv T_{YY}/(e^2f^4)=-\frac{2(w-y_0^2)]\,
[1+\mathcal{A}(w)]\,(1-\mathcal{F})}
{w^2[1+C_1/\sqrt{w}-2\widetilde{\eta}\, 
[1-\mathcal{A}(w)](1-\mathcal{F})]}\,.
\end{equation}
For $\mathcal{F}=-1$, the component $t_{yy}$
vanishes if $w=-1+4\gamma/3$. The component
$t_{tt}$ vanishes and the component $t_{yy}$ develops a pole in $w$ if 
\begin{equation}
 \sqrt{w}=-\frac{C_1}{2(1-4\widetilde \eta)}
\left(1\mp\sqrt{1-\frac{16\, \widetilde{\eta}\, (1-4\widetilde{\eta})}{C_1^2}
\left(1-4\gamma/3\right)}\right)\,.
\end{equation}
As $\sqrt{w}\ge y_0$, these poles can be avoided 
by taking the size of the defect, $y_0$, to be sufficiently large.

The dimensionless angular pressure components are given by   
\bsubeqs\beqa
t_{zz}
&\equiv&
T_{ZZ}/(e^2 f^2)
= \frac{(1-\mathcal{F})^2 \, (1-4\gamma/3)}{w}\,,
\\[2mm]
t_{xx}
&\equiv&
T_{XX}/(e^2 f^2)
= 
(\sin\,Z)^2 \; t_{zz}\,.
\eeqa\esubeqs

Three remarks can be made.
First, only the $t_{yy}$ component depends explicitly on $y_0$
(and goes to zero for $w\to y_0^2$) because, for constant $\widetilde F(w)$, 
the other components do not depend on $\widetilde\kappa$. 
Second, the angular pressure components vanish identically for the 
particular value $\gamma=3/4$, whereas they become negative for $\gamma>3/4$.
Third, in terms of the energy density $\rho$ and the pressures $p_n$ 
of an imperfect fluid, the asymptotic behavior of  
the Skyrme field gives $\rho\sim +4/w$, $p_y\sim -4/w$,
and $p_x\sim p_z = O(1/w^2)$, 
so that $\rho+p_y$ vanishes to leading order
(recall that, for a perfect fluid, precisely the combination
$\rho+p$ enters the hydrostatic equilibrium equation~\cite{Weinberg1972};
see also the discussion in Ref.~\cite{VilenkinShellard1986}).

\begin{figure}[t]  
\includegraphics[width=0.8\textwidth]{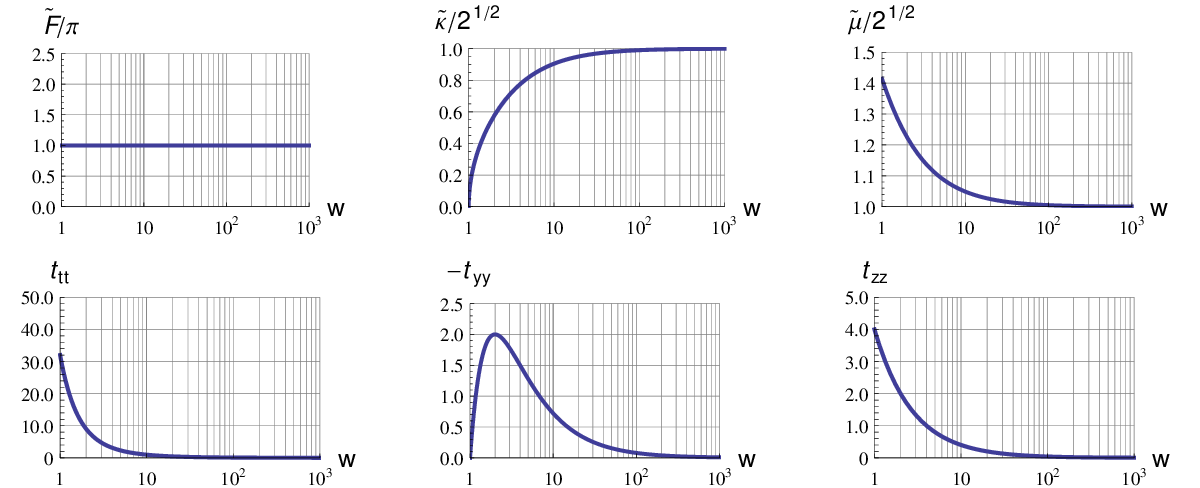}
\caption{\label{fig:EMtensor-gamma-zero}
Top row: functions $\widetilde{F}(w)$, $\widetilde\kappa(w)$, 
and  $\widetilde\mu(w)$ of the nonvacuum solution \eqref{eq:matter-solution}.
Bottom row: corresponding dimensionless energy density $t_{tt}$ 
and pressure components $t_{yy}$ and $t_{zz}$.
The model parameters are $y_0=1$, $\widetilde{\eta}=1/8$,
 and $\gamma=0$.
The solution parameters (from the boundary conditions at the defect core
and the model parameters) are $\mathcal{F}=-1$, $C_{1}=0$, 
and $C_{2}=2$.}   
\vspace*{5mm}
\includegraphics[width=0.8\textwidth]{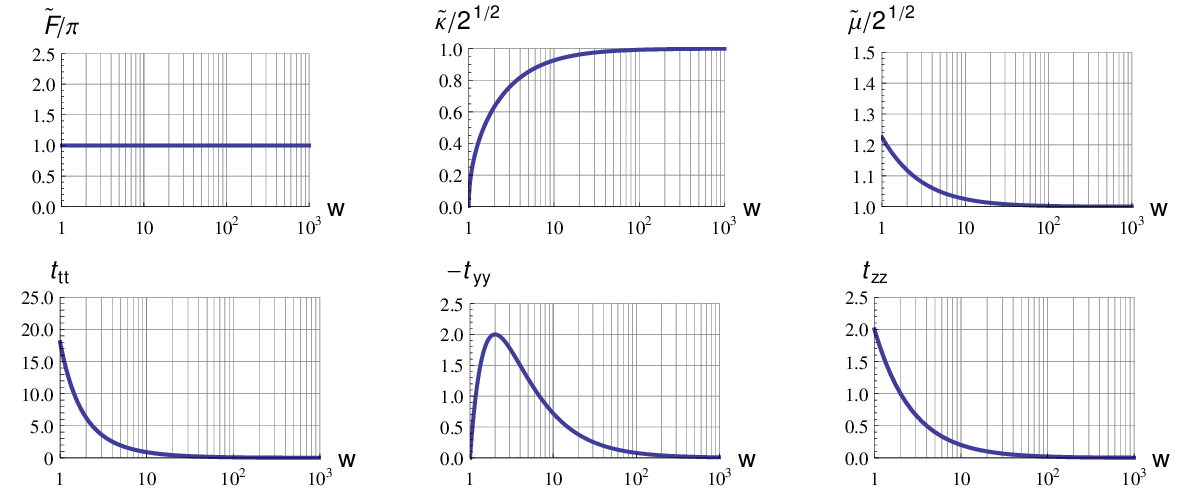}
    \caption{\label{fig:EMtensor-gamma-three-eights}
Same as Fig.~\ref{fig:EMtensor-gamma-zero}, but with $\gamma=3/8$.}
\vspace*{5mm}
\includegraphics[width=0.8\textwidth]{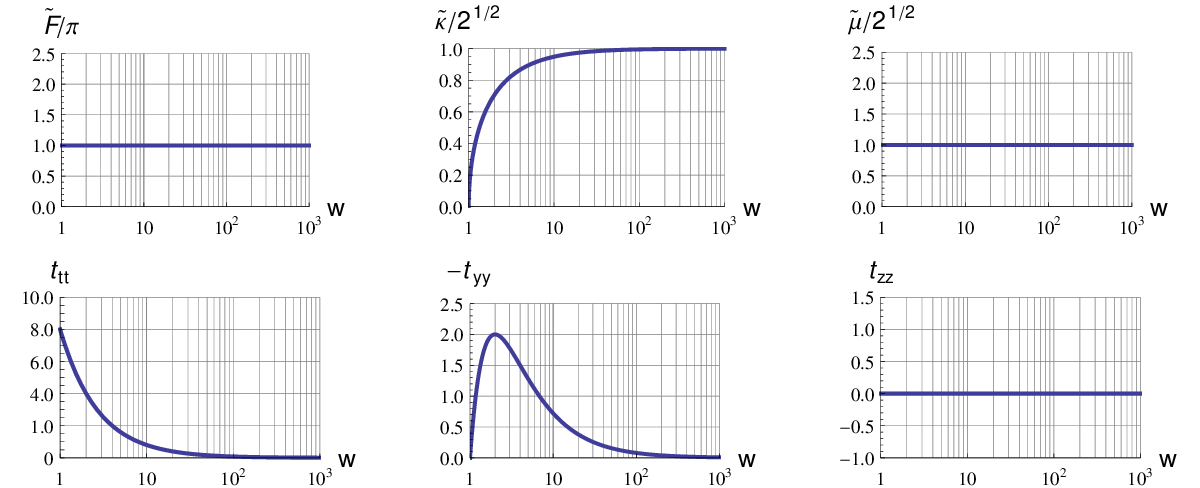}
    \caption{\label{fig:EMtensor-gamma-three-quarter}
Same as Fig.~\ref{fig:EMtensor-gamma-zero}, but with $\gamma=3/4$.}
\end{figure}

As a concrete  example, consider the energy-momentum densities 
for the $C_{1}=0$ case discussed in 
the second paragraph of Sec.~\ref{subsec:Asymptotics-boundary-conditions}.
Figure \ref{fig:EMtensor-gamma-zero} 
shows the components of the energy-momentum tensor 
for a particular choice of parameters, including $\gamma=0$.
Figures  \ref{fig:EMtensor-gamma-three-eights} 
and \ref{fig:EMtensor-gamma-three-quarter}
use the same parameters
but now with $\gamma=3/8$ and $\gamma=3/4$, respectively.
Observe that the value of the
central energy density $t_{tt}(y_0^2)$
decreases with increasing $\gamma$.

\section{Discussion}  
\label{sec:Discussion}

The exact solution from Sec.~\ref{sec:Nonvacuum-solution} 
has a constant nonvanishing Skyrme function and diverging total energy.
Most likely, this particular nonvacuum solution is unstable
and decays to a Skyrmion-type defect solution
by emitting outgoing waves of scalars. It remains to
obtain this final stable Skyrmion-type defect solution
with boundary conditions $\widetilde{F}(y_0^2)=\pi$ and 
$\lim_{w\to\infty}\widetilde{F}(w)=0$.

Note that, strictly speaking, the $SO(3)$ scalar field
is absent in the standard model of elementary-particle physics. 
But the nonlinear
sigma model resurfaces if the gauge fields
are eliminated and the Higgs-field modulus is frozen.
Hence, the toy model of matter fields considered in
the present article is not completely removed from
realistic physics. As such, there may even be
a connection with ideas linking the
quantum structure of spacetime
(having an energy scale $E_\text{Planck}\equiv \sqrt{\hbar\, c^5/G_{N}}$
and a length scale $\hbar c/E_\text{Planck}$)
to the Higgs-boson and top-quark masses~\cite{Klinkhamer2013-JETPL}.
We refrain from making even wilder speculations regarding phenomenology.
In fact, the main focus of this article is purely theoretical, 
namely, finding a nonsingular spacetime-defect solution
with nontrivial matter fields.

\section*{\hspace*{-4.5mm}ACKNOWLEDGMENTS}
\vspace*{-0mm}\noindent
The authors thank the referee for helpful remarks. 
This work has been supported, in part, by the German Research Foundation
(Deutsche Forschungsgemeinschaft, DFG) under Grant No. KL 1103/2-1.

\begin{appendix}
\section{ODES}
\label{app:ODEs}

In this appendix, the relevant ordinary differential equations
(ODEs) are given in detail,
whereas they were only given in abbreviated form in \eqref{eq:final-ODEs-short}.
It turns out to be helpful to define the following dimensionless
auxiliaries:
\bsubeqs
\begin{eqnarray}
u_1&\equiv&\left(1-\frac{4}{3}\,\gamma \right)
\sin ^2\left(\widetilde{F}(w)/2\right)\,, \\[2mm]
u_2&\equiv&\left(1-\frac{2}{3}\,\gamma \right) \sin ^2
\left(\widetilde{F}(w)/2\right)  + \frac{w}{4}\,,
\end{eqnarray}
\esubeqs
dropping the argument $w$ of $u_1$ and $u_2$.

The reduced Einstein equations can now be written in the form
$G_{\;\;Y}^{Y}=8\pi G_{N} \, T_{\;\;Y}^{Y}$, 
$G_{\;\;Y}^{Y}-G_{\;\;T}^{T}=8\pi G_{N} \, (T_{\;\;Y}^{Y}-T_{\;\;T}^{T})$, and
$G_{\;\;Z}^{Z}=8\pi G_{N} \, T_{\;\;Z}^{Z}$
[it is also found that $G_{\;\;X}^{X}=G_{\;\;Z}^{Z}$ 
and $T_{\;\;X}^{X}=T_{\;\;Z}^{Z}$]. In terms of
dimensionless quantities, these reduced Einstein equations are
\bsubeqs
\begin{eqnarray}
\hspace*{-6mm}&&
e^{-2 \widetilde{\lambda}} \left(1+2 \sqrt{w-y_0^2} \,\widetilde{\nu}'\right)-1
=
\nonumber\\[2mm]
\hspace*{-6mm}&&
\widetilde{\eta} \left(4 \sin^2(\widetilde{F}/2) 
\left(\frac{u_1}{w-y_0^2}+1\right)-2 e^{-2 \lambda}
\left(u_2-\frac{y_0^2}{4}\right) \widetilde{F}'^2+\frac{\gamma}{4}\, 
   e^{-4 \lambda}\, \left(w-y_0^2\right) \widetilde{F}'^4\right)
\,,
   \\[2mm]
\hspace*{-6mm}&&
2 \sqrt{w-y_0^2} \left(\widetilde{\lambda}'+\widetilde{\nu}'\right)
=
\nonumber\\[2mm]
\hspace*{-6mm}&&
\widetilde{\eta} \left(
-\left(4 u_2-y_0^2\right) \widetilde{F}'^2
+\frac{\gamma}{3}\, e^{-2 \lambda}\, \left(w-y_0^2\right)\widetilde{F}'^4
\right)
\,,\\[2mm]
\hspace*{-6mm}&&
e^{-2 \widetilde{\lambda}}\, \sqrt{w-y_0^2}\,
\left(\left(\widetilde{\nu}'-\widetilde{\lambda}'\right)
\left(1+\sqrt{w-y_0^2}\, \widetilde{\nu}'\right)+\sqrt{w-y_0^2}\, \widetilde{\nu}''\right)
=\nonumber\\[2mm]
\hspace*{-6mm}&&
\widetilde{\eta} \left(
  -\frac{4 u_1\, \sin^2(\widetilde{F}/2)}{w-y_0^2}
+\frac{1}{2}\, e^{-2 \lambda}\, \left(w-y_0^2\right) \widetilde{F}'^2
 \left(1-\frac{\gamma}{6}\,e^{-2 \lambda}\, \widetilde{F}'^2\right)
\right)
\,.
\end{eqnarray}
\esubeqs
The reduced matter field equation is 
\begin{eqnarray}
  0&=&  \frac{e^{2 \lambda}}{w}\, (\sin \widetilde F) \, 
\left(u_1+\frac{w}{2}\right)
  - \left(\left(w-y_0^2\right) 
\left(1+4 \, u_2\,  (\widetilde\nu '-\widetilde\lambda ')\right)
  +2 u_2\right)\widetilde{F}'
\nonumber\\[2mm]
\hspace*{-6mm}
&&
-\left(1-\frac{2}{3}\gamma\right) \left(w-y_0^2\right) 
(\sin \widetilde F) \,\widetilde{F}'^2
\nonumber\\[2mm]
\hspace*{-6mm}
&&
-\frac{2}{3}\, \gamma\,  e^{-2 \lambda}\,  \left(w-y_0^2\right) 
\left(2 w \left(w-y_0^2\right) 
\left(3 \widetilde\lambda '-\widetilde\nu '\right)-5 w+2 y_0^2\right)\widetilde{F}'^3
\nonumber\\[2mm]
\hspace*{-6mm}
&& 
-4  \left(w-y_0^2\right)
   \left(u_2-\gamma\,  e^{-2 \lambda}\, w \,\left(w-y_0^2\right)\,
    \widetilde{F}'^2\right)\,\widetilde{F}'' \,.
\end{eqnarray}

Using $\widetilde{\kappa}(w)$  and  $\widetilde\mu(w)$   
as defined in \eqref{eq:kappatilde-mutilde}
and solving for $\widetilde{\kappa}'$, $\widetilde{\mu}'$,
and $\widetilde{F}''$, one obtains the final ODEs     
\bsubeqs\label{eq:final-ODEs}  
\begin{eqnarray}\label{eq:final-ODEs-a} 
   \widetilde{\kappa}'
   &=&
   \frac{1 }{4 w\, \widetilde{\kappa}
   \left(w-y_0^2\right)}
   \left(\widetilde{\kappa}^4
   \left[4 \widetilde{\eta}\, \left(u_1+w\right)\,
\sin^2(\widetilde{F}/2)-w\right]+\widetilde{\kappa}^2
   \left(w+y_0^2\right)
   \right.
   \nonumber\\
&& \left.
+
4 \widetilde{\eta}\,\widetilde{\kappa}^2\, \left(w-y_0^2\right)
\left[-\left(1-2\gamma/3\right) w\, \cos (\widetilde F)
\right.\right.\nonumber\\
&&+\left.\left.\left(\left(w-y_0^2\right)^3
+y_0^2\right) \left(1-2\gamma/3+y_0^2\right)
+w \left(w/2-y_0^2\right) \right]\, \widetilde{F}'^2
\right.\nonumber\\
   && \left.  -(4\gamma/3) \, \widetilde{\eta}\, \left(w-y_0^2\right){}^3
   \left[w\, (1+ y_0^4)+y_0^2\, (1-y_0^4)\right]\,  \widetilde{F}'^4\right)
  ,
\eeqa
\beqa\label{eq:final-ODEs-b} 
\widetilde\mu'/\widetilde\mu
&=&
\frac{\widetilde{\kappa}^2 \left[1/4
-\widetilde{\eta} \left(u_1/w+1\right)\,
\sin^2(\widetilde{F}/2)\right]}{w-y_0^2}
   -\frac{1}{4 w}
\nonumber\\
&&
+2 u_2\, \widetilde{\eta}\, \widetilde{F}'^2
-\frac{\gamma\,  w\, \widetilde{\eta}\, \left(w-y_0^2\right)\,\widetilde{F}'^4}
   {\widetilde{\kappa}^2}
\ ,
\eeqa
\beqa\label{eq:final-ODEs-c} 
   \widetilde{F}''
   &=&
   \frac{\widetilde{\kappa}^2}{\left(w-y_0^2\right)
   \left(u_2\, \widetilde{\kappa}^2-\gamma\,  w\, \left(w-y_0^2\right){}^3
   \,\widetilde{F}'^2\right)}
   \left[
   \frac{\widetilde{\kappa}^2}{8}
   \left(1+\frac{2 u_1}{w}\right) \sin(\widetilde{F})
   \right.\nonumber\\
 &&\left.
   -\frac{1}{2}  \left(u_2\,
   \widetilde{\kappa}^2 \left(1-4 \widetilde{\eta}\, \left(u_1/w+1\right)\,
  \sin^2(\widetilde{F}/2)\right)
   +\frac{1}{2} \left(w-y_0^2\right)\right)\,\widetilde{F}'\right.\nonumber\\
 &&-\left.\frac{1}{4} \left(1-2\gamma/3\right)
   \left(w-y_0^2\right)  \sin (\widetilde{F})\, \widetilde{F}'^2
  \right.\nonumber\\
 &&\left.
   -\gamma  \left(w-y_0^2\right)
  \left(\frac{4}{3}\, \widetilde{\eta}\, \left(u_1+w\right) 
\sin ^2(\widetilde{F}/2)
   -\frac{w-y_0^2}{2\, \widetilde{\kappa}^2}-\frac{w}{3}\right)\, \widetilde{F}'^3
\right.\nonumber\\
 &&\left.
    -\frac{2\; \gamma\;   \widetilde{\eta}\; w\left(w-y_0^2\right)^4  u_2}
      {3\, \widetilde{\kappa}^2}\,\widetilde{F}'^5
   \right] \,, 
\end{eqnarray}
\esubeqs
where terms with equal powers of $\widetilde{F}'$ have been grouped together.

\section{VACUUM-SOLUTION GEODESICS} 
\label{app:Geodesics}

For the vacuum solution of Sec.~\ref{sec:Vacuum-solution}, the
nontrivial Christoffel symbols are
\begin{subequations}    \begin{eqnarray}
\frac{1}{e f}\,\Gamma^T_{\ T Y}&=&
\frac{\sqrt{w-y_0^2}\;\, \ell/2}{w^{3/2}-\ell\,w}\,,
\\[1mm]
\frac{1}{e f}\,\Gamma^Y_{\ T T}&=&\frac{
\left(\sqrt{w}-\ell \right)\;\ell/2}{ w \sqrt{w-y_0^2}}\,,
\\[1mm]
\frac{1}{e f}\,\Gamma^Y_{\ Y Y}
&=&
\frac{\sqrt{w}\;\, y_0^2-\left(w+y_0^2\right)\,\ell/2}
     {w \left(\sqrt{w}-\ell \right)\sqrt{w-y_0^2}}\,.
\end{eqnarray}
\end{subequations}
The last two Christoffel symbols are seen to diverge at the defect core
($w=y_0^2$),
but the particle motion can still be regular, as we will now show
by a simple example [using dimensionless Christoffel symbols
$\gamma^{a}_{bc}\equiv (ef)^{-1}\,\Gamma^{A}_{BC}$].

Consider the geodesic equation for a particle moving solely in the $y$ direction 
and denote the nonvanishing dimensionless velocity components
$u^t=dt/d\lambda$ and $u^y=dy/d\lambda$,
where $\lambda$ is a dimensionless parameter and $c=1$.
The particle motion is then given by the following two equations:
\begin{subequations}\label{eq:godesics-ut-uy}
\begin{eqnarray}
0&=&\frac{d u^t}{d\lambda}+2\,\gamma^t_{\ t y}\,u^t u^y\, ,
\\[1mm]
0&=&\frac{d u^y}{d\lambda}+\gamma^y_{\ t t}\,u^t u^t
+\gamma^y_{\ y y}\,u^y u^y\, .
\end{eqnarray}
\end{subequations}  The first equation can be solved for $u^t$:
\begin{equation}
u^t=\left(1-\frac{\ell}{\sqrt{w}}\right)^{-1}\, .
\end{equation}
Inserting this $u^t$ into the second equation gives
\begin{equation}\label{eq:godesics-uy-tmp2}
  \frac{d^2 y}{d\lambda^2}
+\frac{-y^2\,\ell+2 y_0^2 \left(-\ell
   +\sqrt{y_0^2+y^2}\right)}
   {2 y \left(y^2+y_0^2\right) \left(-\ell
+\sqrt{y_0^2+y^2}\right)}\left(\frac{d y}{d\lambda}\right)^2
   =\frac{\ell}{2 y \left(\ell-\sqrt{y_0^2+y^2}\right)}\; .
\end{equation}
Finally, replace in \eqref{eq:godesics-uy-tmp2}
\begin{equation}
\frac{dy}{d\lambda}=\frac{1}{2 y}\frac{dy^2}{d\lambda}\ , \
\frac{d^2y}{d\lambda^2}=\frac{1}{2 y}\frac{d^2y^2}{d\lambda^2}-\frac{1}{4y^3}
\left(\frac{dy^2}{d\lambda}\right)^2
\end{equation}
to obtain
\begin{equation}
\frac{d^2y^2}{d\lambda^2}-\frac{1}{4\,(y_0^2+y^2)}\left(1+\frac{\sqrt{y_0^2+y^2}}
{\sqrt{y_0^2+y^2}-\ell}\right)
\left(\frac{dy^2}{d\lambda}\right)^2
=\frac{-\ell}{\sqrt{y_0^2+y^2}-\ell}\; ,
\end{equation}
which remains finite along the trajectory as long as $\ell<y_0$.

For the special case $\ell=0$ (flat spacetime with a ``hole''), it is
possible to obtain explicit solutions of
\eqref{eq:godesics-ut-uy}.  Up to arbitrary time shifts,
the radial geodesics are given in terms of two real constants $A$ and $B$, 
with positive $B$:  
\bsubeqs\label{eq:radial-geodesics}
\beqa\label{eq:radial-geodesics-zero-vel}
y(t) &=& A\,y_0\,,
\\[2mm]\label{eq:radial-geodesics-nonzero-vel}
y(t) &=& \left\{\begin{array}{ll}
                \pm\,y_0\,\sqrt{(B\, t)^2+2\, B\, t}&\quad\text{for}\quad  t \geq 0\,,\\
                \mp\,y_0\,\sqrt{(B\, t)^2-2\, B\, t}&\quad\text{for}\quad  t < 0\,,
              \end{array}\right.
\eeqa
\esubeqs
where the upper entries before $y_0$
on the right-hand side of \eqref{eq:radial-geodesics-nonzero-vel}
correspond to a positive asymptotic velocity
and lower entries correspond to a negative asymptotic velocity.
Observe that $y^2$ from the second solution is nondifferentiable at $t=0$.

\end{appendix}


\begin{thebibliography}{99}
\bibitem{Endnote:invariants}
Specifically, the 14 curvature invariants calculated are:
$R$\,, $R_{\mu\nu}\, R^{\mu\nu}$\,, $R_{\mu\nu\rho\sigma}\, R^{\mu\nu\rho\sigma}$\,, 
$R^{\mu\nu}\, R^\rho_{\phantom{\beta}\mu}\, R_{\nu\rho}$\,, 
$R^{\mu\nu\rho\sigma}\, R_{\mu\rho}\, R_{\nu\sigma}$\,, 
$R^{\mu\nu\rho\sigma}\, R^\tau_{\phantom{\beta}\mu}\, R_{\nu\rho\sigma\tau}$\,, 
$R^{\mu\nu\rho\sigma}\, R^{\tau\upsilon}_{\phantom{\beta}\phantom{\beta}\mu\nu}\, R_{\rho\sigma \tau\upsilon}$\,, 
$R^{\mu\nu\rho\sigma}\, R^{\tau\phantom{\beta} \upsilon}_{\phantom{\beta}\mu\phantom{\beta} \rho}\, R_{\nu \tau\sigma \upsilon}$\,, 
$R^{\mu\nu\rho\sigma}\, R^{\tau\phantom{\beta} \upsilon}_{\phantom{\beta}\mu\phantom{\beta} \rho}\, R_{\nu \upsilon\sigma \tau}$\,, 
$R^{\mu\nu;\rho}\, R_{\mu\nu;\rho}$\,, 
$R^{\mu\nu;\rho}\, R_{\mu\rho;\nu}$\,, 
$R^{\mu\nu}_{\phantom{\beta}\phantom{\beta};\mu}\, R^\rho_{\phantom{\beta}\nu;\rho}$\,, 
$R^{\mu\nu\rho\sigma;\tau}\, R_{\mu\nu\rho\sigma;\tau}$\,, and
$R^{\mu\nu\rho\sigma}_{\phantom{\beta}\phantom{\beta}\phantom{\beta}\phantom{\beta};\mu}\, 
R^\tau_{\phantom{\beta}\nu\rho\sigma;\tau}$\,.
\end{thebibliography}
\end{document}